\documentclass[preprint,preprintnumbers,amsmath,amssymb]{revtex4}
\usepackage{graphicx}
\usepackage{epstopdf}
\usepackage{dcolumn}
\usepackage{bm}
\usepackage{color}
\usepackage{array,multirow,graphicx}

\begin{document}

\title{Intercalation of O$_2$ and N$_2$ in the Graphene/Ni Interfaces of Different Morphology}

\author{Changbao Zhao,$^{1,2}$ Jiayi Li,$^{1,2}$ Jiuxiang Dai,$^{1,3}$ Elena Voloshina,$^{4,}$\footnote{Corresponding author. E-mail: voloshina@shu.edu.cn} Yuriy Dedkov,$^{4,}$\footnote{Corresponding author. E-mail: dedkov@shu.edu.cn} Yi Cui$^{1,}$\footnote{Corresponding author. E-mail: ycui2015@sinano.ac.cn}}

\affiliation{$^1$Vacuum Interconnected Nanotech Workstation, Suzhou Institute of Nano-Tech and Nano-Bionics, Chinese Academy of Sciences, 215123 Suzhou, China}
\affiliation{$^2$Nano Science and Technology Institute, University of Science and Technology of China, 230026 Hefei, China}
\affiliation{$^3$Department of Chemistry, Shanghai University, 200444 Shanghai, China}
\affiliation{$^4$Department of Physics, Shanghai University, 200444 Shanghai, China}

\date{\today}

\begin{abstract}

Near-ambient pressure XPS and STM experiments are performed to study the intercalation of oxygen and nitrogen at different partial gas pressures and different temperatures in the graphene/Ni/Ir(111) system of different morphologies. We performed detailed experiments on the investigation of the chemical state and topography of graphene, before and after gas intercalation, depending on the amount of pre-intercalated Ni in graphene/Ir(111). It is found that only oxygen can be intercalated under graphene in all considered cases, indicating the role of the intra-molecular bonding strength and possibility of gas molecules dissociation on different metallic surfaces on the principal possibility and on the mechanism of intercalation of different species under graphene.\\

This document is the unedited author's version of a Submitted Work that was subsequently accepted for publication in J. Phys. Chem. C., copyright $\copyright$ American Chemical Society after peer review. To access the final edited and published work see doi: 10.1021/acs.jpcc.9b01407.

\end{abstract}

\maketitle

\section{Introduction}
\label{intr}

The epitaxial growth of different 2D materials, like graphene (gr), h-BN, MoS$_2$, etc. on different substrates~\cite{Dedkov:2015kp,Auwarter:2018kf,Li:2016hn} gives an opportunity to study the influence of low-dimensionality on the reactivity of the studied systems, i.\,e. the effects of the so-called 2D confined catalysis~\cite{Deng:2016ch,Fu:2016gt}. These effects include (but not limited to) intercalation of different species in the space between 2D material and support, catalysis under 2D cover, growth under 2D cover of adlayers with the artificial crystallographic and electronic structures, solution intercalation, etc. Here as examples one can consider intercalation of different elements (metals and non-metals)~\cite{Dedkov:2015kp,Batzill:2012,Dedkov:2001,Shikin:2000,Emtsev:2011fo,Verbitskiy:2015kq}, intercalation of molecules (like CO, H$_2$O, C$_{60}$)~\cite{Feng:2012il,Politano:2016jb,Gupta:2014he,Shikin:2000a,Monazami:2015gg,Zhang:2015hm,Emmez:2014cl,Gurarslan:2014bx}, stabilization of the pseudomorphic growth of metals on different substrates~\cite{Dedkov:2008e,Pacile:2013jc,Decker:2013ch}, CO oxidation under graphene or h-BN on Pt(111)~\cite{Zhang:2015hm,Yao:2014hy} (see for review Refs.~\citenum{Deng:2016ch,Fu:2016gt} and references therein).

Previous experimental and theoretical works on the confined catalysis dealt with the epitaxial 2D layers grown on the bulk support and the catalytic activity of such systems was taken as is without attempts to modify their properties via change of system morphology or its electronic properties. Here, one can assume that, e.\,g. the intercalation of guest-metal atoms in the epitaxial gr/metal or gr/semiconductor interface might tune the properties of these interfaces. One can think, for example, on the creation of additional reaction centres from atoms or clusters of guest-metal and/or on the formation of the new artificial gr/guest-metal interface, where the system morphology and the low-dimensionality of the intercalated metal will influence the catalytic properties of the studied systems. For example, considering the artificial gr/$n$\,ML-Ni/Ir(111) or gr/$n$\,ML-Co/Ir(111) systems, which can be formed via intercalation of Ni or Co, respectively, in gr/Ir(111)~\cite{Pacile:2013jc,Vu:2016ei}, several factors, like gr-metal lattice mismatch, graphene morphology, quantum size effects in the electronic structure of the Ni or Co layer, might influence the catalytic activity of such gr/Ni or gr/Co interfaces.

Here we present the experiments on the oxygen and nitrogen intercalation in gr/$n$\,ML-Ni/Ir(111), where thickness of the Ni layer is varied from sub-monolayer to bulk-like thick film. Our near-ambient pressure x-ray photoelectron spectroscopy (XPS) and scanning tunnelling microscopy (STM) studies allowed to discriminate between different factors, like system morphology and the electronic structure of thin Ni films, which govern the oxygen intercalation and formation of thin NiO layer under graphene. Surprisingly, our attempts on the N$_2$ intercalation in the gr/Ni/Ir(111) interface in a wide range of the gas pressure and substrate temperature did not succeed, indicating the stronger N-N bond compared to the O-O one and influence of the catalytically active metal substrates under graphene on the molecules decomposition during intercalation. Our experiments demonstrate the possibility to tune the catalytic properties of the gr/metal system in the confined catalysis processes and will stimulate further studies in this research field.

\section{Results and discussions}
\label{ResDisc}

\subsection{Oxygen intercalation in gr/Ir(111): NAP-XPS and STM} 

\begin{figure}[t]
	\centering
	\includegraphics[width=0.7\columnwidth]{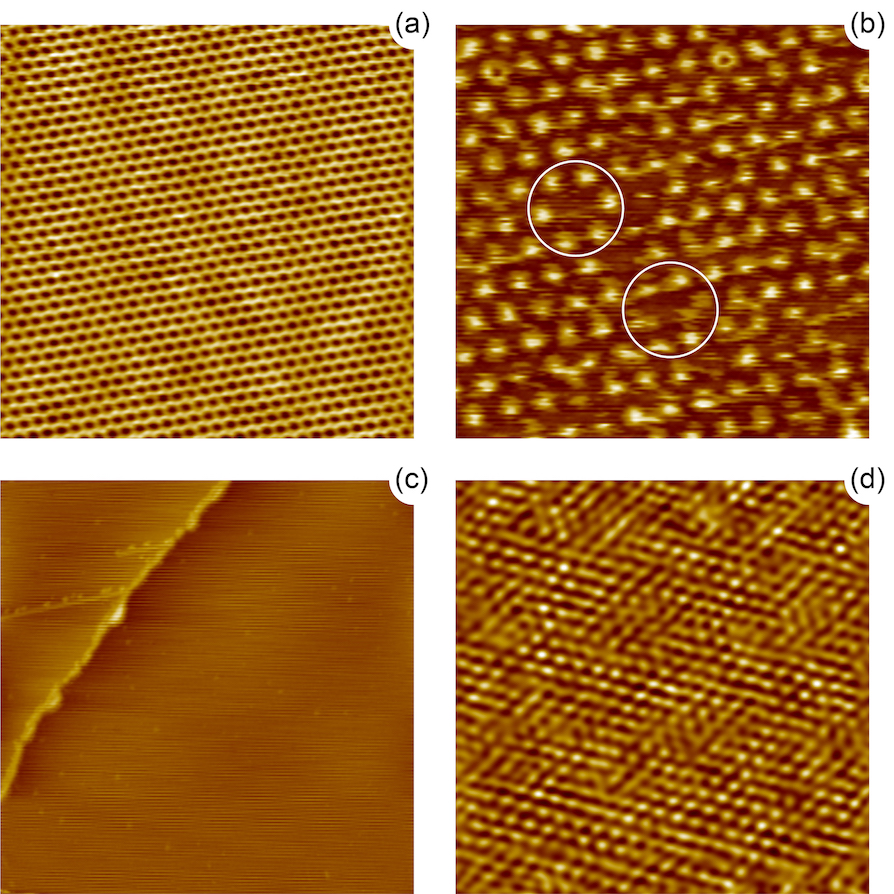}
	\caption{STM images of gr/Ir(111) before and after oxygen intercalation: (a) gr/Ir(111), $75\times75\,\mathrm{nm}^2$, $U_T=+0.17\,\mathrm{V}$, $I_T=0.68\,\mathrm{nA}$, (b) after exposure of gr/Ir(111) to O$_2$ at $0.1$\,mbar and $200^\circ$\,C for $2$\,min (partial intercalation), $30\times30\,\mathrm{nm}^2$, $U_T=+0.58\,\mathrm{V}$, $I_T=0.42\,\mathrm{nA}$, and (c) after exposure of gr/Ir(111) to O$_2$ at $0.1$\,mbar and $200^\circ$\,C for $50$\,min (complete intercalation), $200\times200\,\mathrm{nm}^2$, $U_T=+1.25\,\mathrm{V}$, $I_T=0.11\,\mathrm{nA}$. Atomically resolved image in (d) shows a zoomed area of (c) with imaging parameters: $7\times7\,\mathrm{nm}^2$, $U_T=+0.01\,\mathrm{V}$, $I_T=1\,\mathrm{nA}$. Circles in (b) mark the high-symmetry position of gr/Ir(111) which disappear after $2$\,min exposure this system to oxygen.}
	\label{grOIr_STM}
\end{figure}

\begin{figure}[t]
	\centering
	\includegraphics[width=0.7\columnwidth]{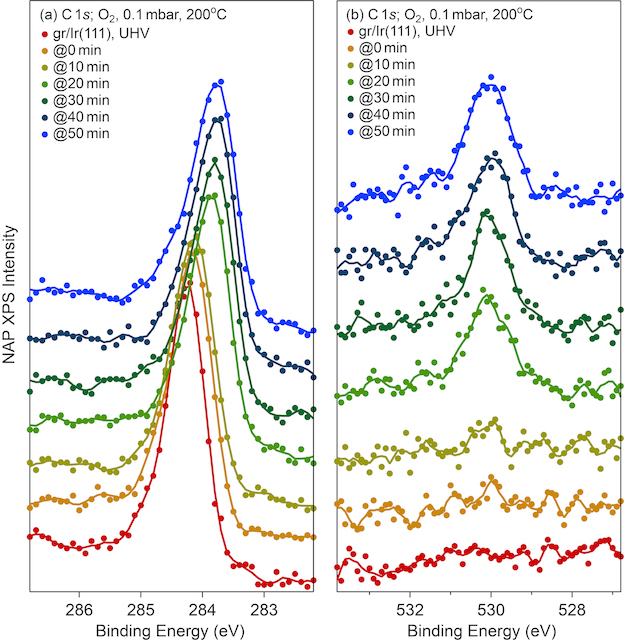}
	\caption{NAP-XPS intensities of (a) C\,$1s$ and (b) O\,$1s$ measured for gr/Ir(111) during its exposure to O$_2$ at $0.1$\,mar and $200^\circ$\,C. The respective exposure time for every spectra is marked in the figure legend. All XPS spectra are shifted for clarity. The smoothed line through the experimental points is shown for every spectra.}
	\label{grOIr111_NAP_XPS}
\end{figure}

We started our studies on the oxygen and nitrogen intercalation in the gr/Ni/Ir(111) system from the well-known example: O$_2$ -- intercalation in the parent gr/Ir(111) interface~\cite{Larciprete:2012aaa,Granas:2012cf}. In our near-ambient-pressure-XPS (NAP-XPS) and NAP-STM studies the Ir(111) substrate was fully covered with a complete graphene layer and the obtained NAP-XPS and STM results are compiled in Figs.~\ref{grOIr_STM} and \ref{grOIr111_NAP_XPS}.

Graphene was grown on Ir(111) via exposing its surface to C$_2$H$_4$ at $10^{-6}$\,mbar and $1400$\,K. This procedure leads to a controllable growth of a high-quality graphene layer which fully covers Ir(111) as deduced from the large-scale STM experiments. In the STM images (Fig.~\ref{grOIr_STM}), gr/Ir(111) displays large fully graphene covered terraces, which are several hundreds nanometer wide and have straight steps following the directions of the graphene moir\'e structure~\cite{Voloshina:2013dq,Dedkov:2014di}. This graphene moir\'e structure is clearly measured in the STM experiments as an additional modulation of the graphene lattice with a periodicity of $25.2$\,\AA\ (Fig.~\ref{grOIr_STM}(a)). It is formed on top of Ir(111) due to the relatively large lattice mismatch of $\approx10$\% between two in-plane lattices. Previously published experimental data~\cite{Hattab:2011ix} indicate the formation of the rotational domains at the temperature which was used during graphene growth on Ir(111) in our experiments. Despite the fact that we cannot fully exclude the formation of such domains during the preparation routine, our present STM results show that number of the corresponding defects (domain boundaries) is very small (if any).

Oxygen intercalation in high-quality gr/Ir(111) was performed at the O$_2$ partial pressure of $0.1$\,mbar and the sample temperatures of $200^\circ$\,C. Fig.~\ref{grOIr_STM}(b) shows the STM image acquired after $2$\,min of oxygen exposure at these conditions and one can see that the moir\'e structure of the gr/Ir(111) system started to become weak. The areas in the STM image shown in panel (b) mark the high-symmetry positions of the gr/Ir(111) structure (namely the ATOP positions) which disappear after described treatment. According to other experiments on the CO intercalation in h-BN/Ru(0001)~\cite{Wu:2018gb} the process of the gas molecules intercalation starts around the defect sites in the 2D layer or/and at the steps and boundaries of the 2D-material/metal structure, where 2D layer becomes flatter after gas molecules penetration under 2D layer. We believe that the similar mechanism can be applied here as it is also supported by the previous studies on the oxygen intercalation in gr/Ir(111)~\cite{Larciprete:2012aaa,Granas:2012cf}. The prolonged exposure of the gr/Ir(111) system to oxygen or/and the slight increase of the temperature used during this procedure leads to the intercalation of oxygen in all areas leading to the flattening of a graphene layer on Ir(111) (Fig.~\ref{grOIr_STM}(c)). After oxygen intercalation in gr/Ir(111) the moir\'e structure of graphene is hardly recognisable in the STM images (Fig.~\ref{grOIr_STM}(c,d)). 

The process of the oxygen intercalation in gr/Ir(111) was further confirmed and studied in the NAP-XPS experiment. Fig.~\ref{grOIr111_NAP_XPS} shows the time evolution of the (a) C\,$1s$ and (b) O\,$1s$ emission lines during the exposure of gr/Ir(111) to O$_2$ at the partial pressure of $0.1$\,mbar and the sample temperature of $200^\circ$\,C. These data show that oxygen gradually intercalates in gr/Ir(111) at these conditions that is manifested as the shift of the C\,$1s$ line to smaller binding energies. After complete oxygen intercalation in gr/Ir(111) the binding energy for C\,$1s$ is $283.7\pm0.1$\,eV compared to the initial value of $284.2\pm0.1$\,eV measured before intercalation for gr/Ir(111). Both values for the binding energy are in good agreement with the previously published data~\cite{Larciprete:2012aaa}. At the same time the intensity of the O\,$1s$ line is monotonically increased with the unchanged binding energy for this line. These data provide a direct spectroscopic evidence that oxygen can diffuse to the gr/Ir(111) interface and adsorb on Ir(111) underneath the graphene overlayer. The shift of the C\,$1s$ line to the smaller binding energies indicates the strong $p$ doping of a graphene layer in the resulting gr/O/Ir(111) systems as confirmed by the previous angle-resolved photoemission (ARPES) data~\cite{Larciprete:2012aaa}. From the XPS data we found that the full-width-at-the-half-maximum (FWHM) of the C\,$1s$ line before oxygen intercalation is $0.82$\,eV and it does not change significantly after oxygen intercalation in gr/Ir(111). From all these results, we conclude that these parameters for the oxygen intercalation (the O$_2$ partial pressure of $0.1$\,mbar and the sample temperature of $200^\circ$) can be used in our further experiments on the studies of the gas intercalations in the gr/Ni/Ir(111) system. At the same time, according to the XPS and STM data, we conclude that oxygen is uniformly intercalated in the gr/Ir(111) system.

\subsection{Oxygen intercalation in gr/$1.6$\,ML-Ni/Ir(111) and gr/Ni(111): NAP-XPS temperature dependence} 

\begin{figure}[t]
	\centering
	\includegraphics[width=0.75\columnwidth]{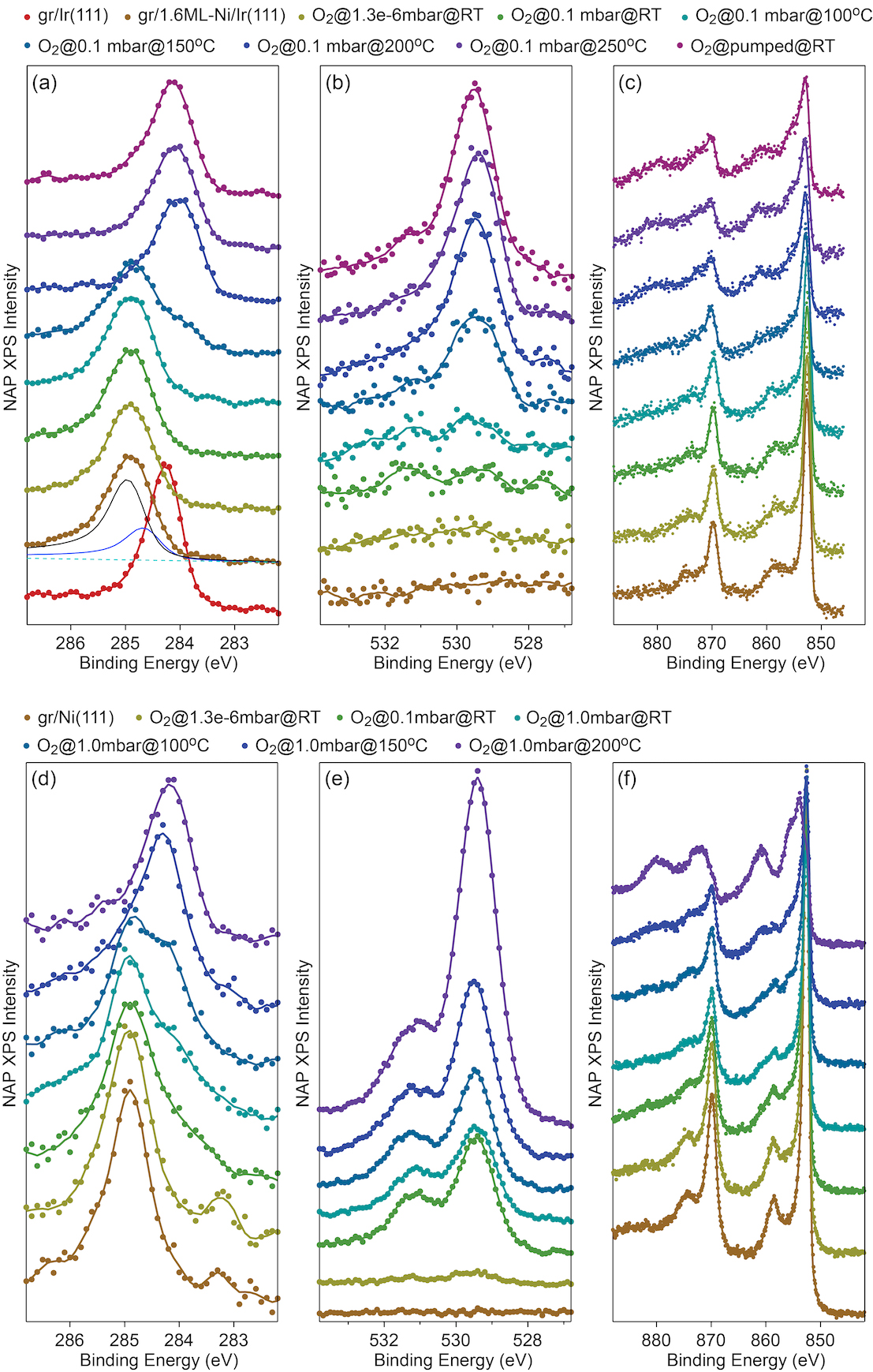}
	\caption{NAP-XPS intensities of (a,d) C\,$1s$, (b,e) O\,$1s$, and (c,f) Ni\,$2p$ measured for gr/$1.6$\,ML-Ni/Ir(111) (top row) and gr/Ni(111) (bottom row) during exposure to O$_2$ at fixed pressure - $0.1$\,mar or $1.0$\,mbar and different sample temperatures. The respective exposure parameters for every spectra are marked in the figure legend. All XPS spectra are shifted for clarity. The smoothed line through the experimental points is shown for every spectra.}
	\label{grONiIr_TempDep}
\end{figure}

In the further experiments, prior to the oxygen or nitrogen intercalation, the gr/Ir(111) interface was modified via intercalation of Ni layer of different thickness. The intercalation of different amounts of Ni in gr/Ir(111) leads to the formation of the gr/Ni/Ir(111) interfaces of different morphologies: from the strongly buckled graphene layer on $1$\,ML-Ni/Ir(111) to flat graphene on thick-Ni/Ir(111)~\cite{Pacile:2013jc}. As tests systems for the gas intercalation we chose gr/$1.6$\,ML-Ni/Ir(111) (strongly buckled graphene) and gr/Ni(111) (flat graphene) systems. The results for the oxygen intercalation in these interfaces are summarized in Fig.~\ref{grONiIr_TempDep} where intensities of the C\,$1s$, O\,$1s$, and Ni\,$2p$ lines are shown as a function of the substrate temperature used in these NAP-XPS measurements.

Before oxygen intercalation, the C\,$1s$ lines for the considered systems have a different shape: for gr/Ni(111) it consists of a single peak located at $284.9\pm0.1$\,eV, whereas for the gr/$1.6$\,ML-Ni/Ir(111) systems this line consists of a main peak at $284.9\pm0.1$\,eV and small shoulder at $284.65\pm0.1$\,eV (see the respective fit of the C\,1s XPS spectrum for gr/$1.6$\,ML-Ni/Ir(111) before oxygen exposure in Fig.~\ref{grONiIr_TempDep}(a)), which is consistent with the previously published results~\cite{Pacile:2013jc}. (The small bump at $283.3\pm0.1$\,eV in the C\,$1s$ spectrum for gr/Ni(111) is due to the small fraction of the Ni$_2$C phase which is rarely avoided during formation of graphene on a Ni(111) single crystal~\cite{Dedkov:2017jn,Spath:2016db}.) For both systems, the emission line for Ni\,$2p$ represents the spin-orbit split doublet at $852.6\pm0.1$\,eV and $869.9\pm0.1$\,eV as well as the correlation satellite peaks which are located $6$\,eV below every main emission line. 

The exposure of both systems to O$_2$ at the partial pressure of $0.1$\,mbar and different samples temperatures indicates the existence of different processes which define the oxygen intercalation under graphene as well as the oxidation of the underlying Ni. For the first case of the gr/$1.6$\,ML-Ni/Ir(111) system, the exposure to O$_2$ at room temperature does not lead to any changes of the C\,$1s$ and Ni\,$2p$ emission lines (Fig.~\ref{grONiIr_TempDep}(a,c)) as was also previously observed for other graphene-based intercalation systems~\cite{Dedkov:2008e,deLima:2014dm,deCamposFerreira:2018hh}. The similar treatment of the gr/Ni(111) interface leads to the partial intercalation of oxygen already at room temperature as can be visible from the appearance of the shoulder on the low binding energy side of the C\,$1s$ peak (third spectrum from the bottom in Fig.~\ref{grONiIr_TempDep}(d)). These conclusions are also supported by the absence of the O\,$1s$ emission for these conditions in case of gr/$1.6$\,ML-Ni/Ir(111) (Fig.~\ref{grONiIr_TempDep}(b)) and its appearance in case of gr/Ni(111) (Fig.~\ref{grONiIr_TempDep}(e)). The same is valid for Ni\,$2p$ where the start of the nickel oxidation is detected for gr/Ni(111) at $0.1$\,mbar and room temperature. From these data we can conclude that the quality and the morphology of a graphene layer for gr/$1.6$\,ML-Ni/Ir(111) is more uniform compared to gr/Ni(111) and that number of defects in graphene is much smaller for the former system. This leads to the effective protection of the underlying intercalated layer for gr/Ni/Ir. One of the reasons which might promote the intercalation of oxygen under graphene on Ni(111) single crystal is the unavoidable formation of the Ni$_2$C phase which always formed during sample preparation. These phase was also observed in our previous and present experiments as confirmed by STM and XPS.

Further increase of the sample temperature above the room temperature or/and the partial oxygen pressure for both considered systems -- gr/$1.6$\,ML-Ni/Ir(111) and gr/Ni(111) -- leads to the effective intercalation of oxygen under graphene and to the oxidation of Ni. In the end of this experiment the binding energy of C\,$1s$ is $284.1$\,eV indicating the complete decoupling of graphene from the substrate and the formation of the gr/NiO interface~\cite{Dedkov:2017jn}. It is interesting to note the difference for two systems with regard to the shape of the O\,$1s$ peak. In the end of the intercalation process it consists of a single component at $529.5\pm0.1$\,eV with a small shoulder $531.3\pm0.1$\,eV for gr/$1.6$\,ML-Ni/Ir(111) and consists of two clear peaks at $529.4\pm0.1$\,eV and $531.2\pm0.1$\,eV for gr/Ni(111). The low binding energy component in both cases can be clearly assigned to the emission from O$^{2-}$ in NiO~\cite{Lorenz:2000cx,Rettew:2011kd}. The second one might be due to the presence of OH$^{-}$ groups on the surface or can be connected with defects in the oxide layer~\cite{Rettew:2011kd,Tyuliev:1999je,Roberts:1984du}. Taking into account the fact that before oxygen intercalation the gr/Ni(111) sample contains some fraction of Ni$_2$C we can conclude that the second peak is dominated by the emission from the oxygen atoms associated with the defects in the nickel oxide layer.

\subsection{Oxygen intercalation in gr/$n$\,ML-Ni/Ir(111): STM morphology} 

\begin{figure}[t]
	\centering
	\includegraphics[width=0.7\columnwidth]{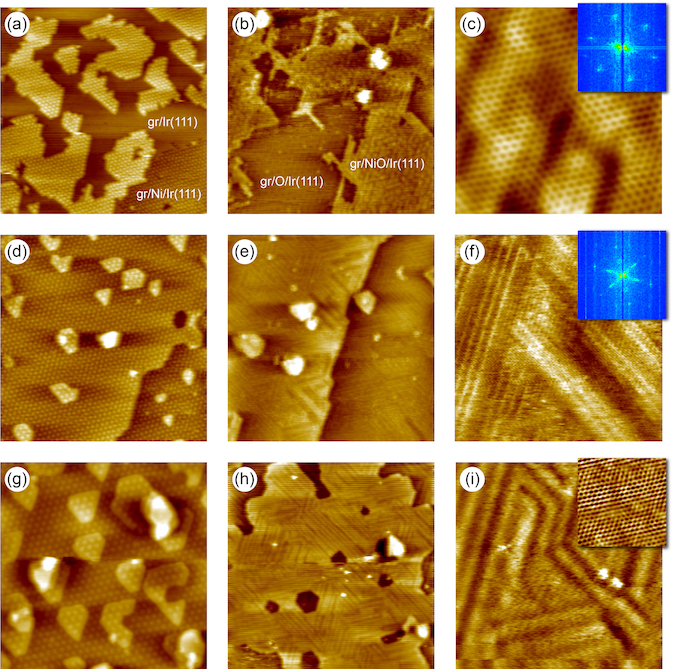}
	\caption{STM images before (left column) and after (middle and right columns) oxygen intercalation in gr/$n$\,ML-Ni/Ir(111): (a-c) gr/$0.5$\,ML-Ni/Ir(111), (d-f) gr/$1.2$\,ML-Ni/Ir(111), and (g-i) gr/$1.6$\,ML-Ni/Ir(111). Insets of (c) and (f) show the respective FFT images and inset of (i) shows the small scale atomically resolved zoom this image. Imaging parameters: (a) $170\times170\,\mathrm{nm}^2$, $U_T=+0.31\,\mathrm{V}$, $I_T=0.35\,\mathrm{nA}$; (b) $100\times100\,\mathrm{nm}^2$, $U_T=+0.45\,\mathrm{V}$, $I_T=0.41\,\mathrm{nA}$; (c) $6\times6\,\mathrm{nm}^2$, $U_T=+0.01\,\mathrm{V}$, $I_T=0.4\,\mathrm{nA}$; (d) $105\times105\,\mathrm{nm}^2$, $U_T=+0.42\,\mathrm{V}$, $I_T=0.29\,\mathrm{nA}$; (e) $100\times100\,\mathrm{nm}^2$, $U_T=+0.02\,\mathrm{V}$, $I_T=0.27\,\mathrm{nA}$; (f) $10\times10\,\mathrm{nm}^2$, $U_T=+0.03\,\mathrm{V}$, $I_T=0.49\,\mathrm{nA}$; (g) $75\times75\,\mathrm{nm}^2$, $U_T=+0.3\,\mathrm{V}$, $I_T=0.56\,\mathrm{nA}$; (h) $100\times100\,\mathrm{nm}^2$, $U_T=+0.12\,\mathrm{V}$, $I_T=0.62\,\mathrm{nA}$; (i) $30\times30\,\mathrm{nm}^2$, $U_T=+0.01\,\mathrm{V}$, $I_T=1.05\,\mathrm{nA}$; iset of (i) $10\times10\,\mathrm{nm}^2$, $U_T=+0.004\,\mathrm{V}$, $I_T=0.44\,\mathrm{nA}$. Parameters for the oxygen intercalation: (b,c) $0.1$\,mbar, $150^\circ$\,C, $60$\,min, (e,f) $0.1$\,mbar, $150^\circ$\,C, $110$\,min, (h,i) $0.1$\,mbar, $150^\circ$\,C, $150$\,min.}
	\label{grONiIr_STM}
\end{figure}

The morphology of the gr/$n$\,ML-Ni/Ir(111) system with different amount of intercalated Ni was investigated by means of STM before and after oxygen intercalation. These results are compiled in Fig.~\ref{grONiIr_STM} where data for (a-c) gr/$0.5$\,ML-Ni/Ir(111), (d-f) gr/$1.2$\,ML-Ni/Ir(111), and (g-i) gr/$1.6$\,ML-Ni/Ir(111) are shown (left column: morphology before oxygen intercalation; middle and right columns: morphology after oxygen intercalation). 

STM results obtained before oxygen intercalation demonstrate high quality of the respective gr/$n$\,ML-Ni/Ir(111) systems where Ni layers of different thicknesses were successfully completely intercalated in gr/Ir(111). For gr/$0.5$\,ML-Ni/Ir(111) (Fig.~\ref{grONiIr_STM}(a)) the areas which are partly modified by Ni can be easily recognized due to the increased corrugation of a graphene layer on Ni which is now pseudomorphically arranged on Ir(111)~\cite{Pacile:2013jc}. Here Ni intercalates at different places of this system -- on terraces as well as around step edges. We found that it is possible to intercalate more than $1$\,ML of Ni in gr/Ir(111) (Fig.~\ref{grONiIr_STM}(d,g)), where first layer of Ni forms the pseudomorphic layer at the gr/Ir(111) interface and the excess amount of Ni forms mono-, double-, and triple-ML thick islands under graphene. The moir\'e structure of a graphene layer is preserved only for the double-layer thick Ni islands indicating the lattice relaxation in the Ni layer for thicker layers. In case of a thick Ni layer (more than $20$\,ML, STM data are not shown) between graphene and Ir(111), the different system preparation was used -- $20$\,ML-thick Ni(111) film was grown on Ir(111) and then graphene was grown via CVD as described in Sec.\,``Experimental'', leading to the lattice-matched system in this case, similar to graphene on the bulk Ni(111), but with the extremely low fraction of the Ni$_2$C phase.

The intercalation of oxygen in gr/$n$\,ML-Ni/Ir(111) leads to the drastic changes in morphology due to the formation of NiO at the gr/Ir(111) interface. After oxygen intercalation in gr/$0.5$\,ML-Ni/Ir(111) two areas can be clearly identified -- gr/O/Ir(111) and gr/NiO/Ir(111) (they are marked in Fig.~\ref{grONiIr_STM}(b)). The first one is very similar to the one presented in Fig.~\ref{grOIr_STM}(c,d) with a faint moir\'e structure, whereas for the gr/NiO/Ir(111) patches the corrugation of graphene is much larger due to the formation of the lattice-mismatched NiO/Ir(111) interface which demonstrates the long-range periodicities similar to the one observed for other NiO,CoO/$4d$,$5d$-metal(111) interfaces~\cite{Hagendorf:2006eo,DeSantis:2011fv,Franz:2012kn,Gragnaniello:2013ip}. After intercalation of oxygen in gr/$n$\,ML-Ni/Ir(111) with thicker Ni layer of more than $1$\,ML, the STM images demonstrate stripe-like and maze-like long-range ordered structures for the NiO layer under graphene, which is also supported by fast Fourier transformation (FFT) images (Fig.~\ref{grONiIr_STM}(e-f,h-i)). It can be explained by the in-plane relaxation in the NiO layer as a function of thickness. In all cases the atomic resolution in graphene layer can be easily obtained where a graphene lattice with all 6 carbon atoms is imaged (Fig.~\ref{grONiIr_STM}: (c) and inset of (i)) confirming the effective decoupling of graphene from the underlying NiO layer. The decoupling of graphene from NiO is also supported by the present XPS data as well as by the recent ARPES and density-functional theory (DFT) studies~\cite{Dedkov:2017jn} where $p$-doping of graphene was observed.

\subsection{Oxygen intercalation in gr/$n$\,ML-Ni/Ir(111): NAP-XPS time dependence} 

\begin{figure}[t]
	\centering
	\includegraphics[width=0.8\textwidth]{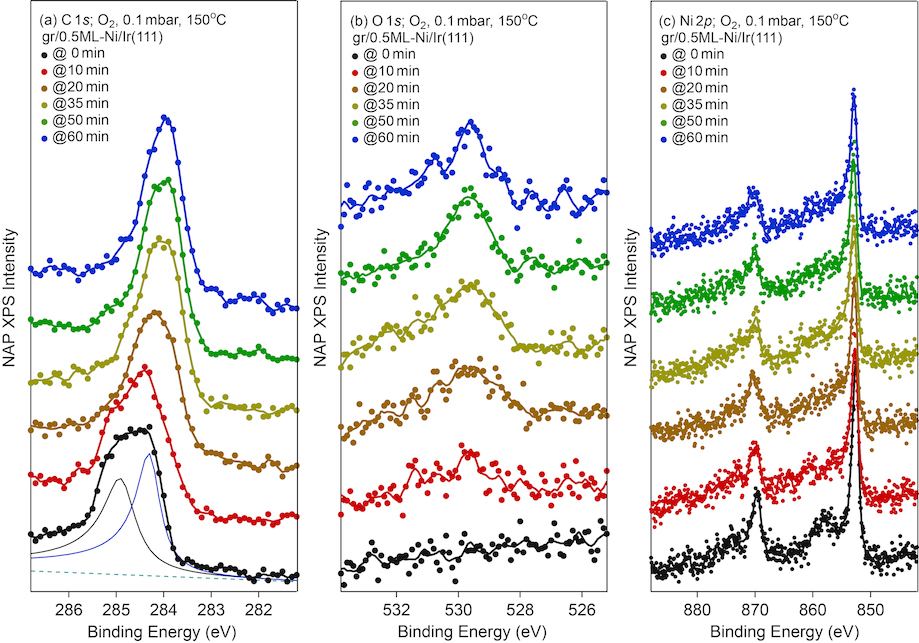}
	\caption{Variation of the NAP-XPS intensities for (a) C\,$1s$, (b) O\,$1s$, and (c) Ni\,$2p$ measured during exposure of the gr/$0.5$\,ML-Ni/Ir(111) system to O$_2$ at fixed pressure ($0.1$\,mar) and fixed sample temperature ($150^\circ$\,C). The respective exposure time for every spectra is marked in the figure legend. All XPS spectra are shifted for clarity. The smoothed line through the experimental points is shown for every spectra.}
	\label{grO05MLNiIr111_NAP_XPS}
\end{figure}

\begin{figure}[t]
	\centering
	\includegraphics[width=0.8\columnwidth]{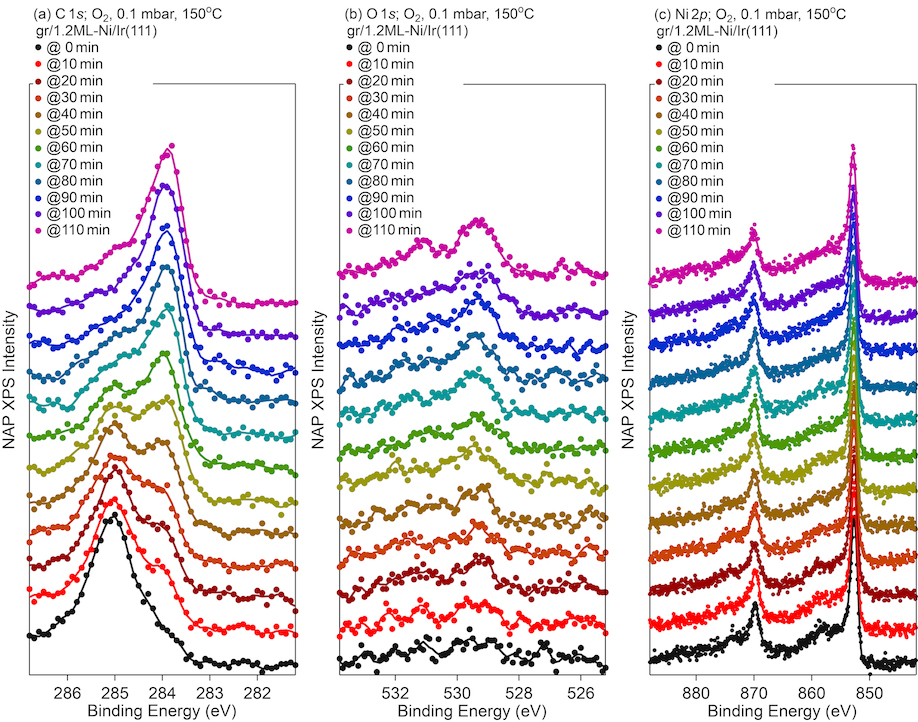}
	\caption{Variation of the NAP-XPS intensities for (a) C\,$1s$, (b) O\,$1s$, and (c) Ni\,$2p$ measured during exposure of the gr/$1.2$\,ML-Ni/Ir(111) system to O$_2$ at fixed pressure ($0.1$\,mar) and fixed sample temperature ($150^\circ$\,C). The respective exposure time for every spectra is marked in the figure legend. All XPS spectra are shifted for clarity. The smoothed line through the experimental points is shown for every spectra.}
	\label{grO12MLNiIr111_NAP_XPS}
\end{figure}

\begin{figure}[t]
	\centering
	\includegraphics[width=0.8\columnwidth]{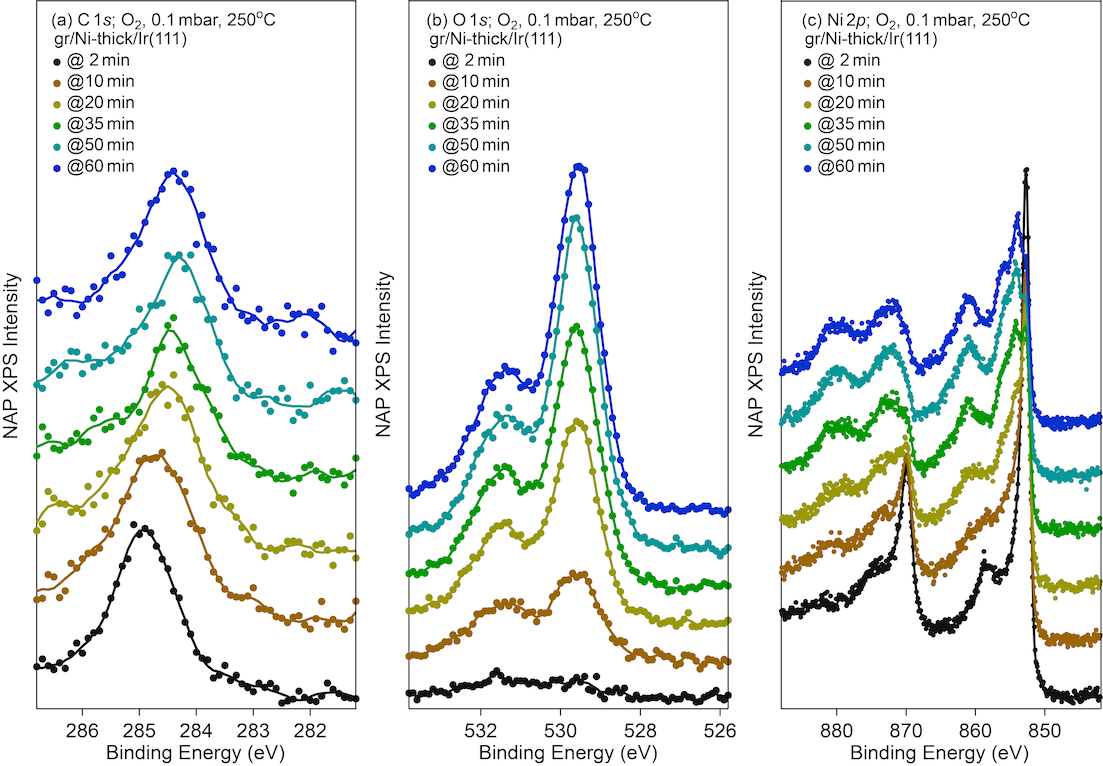}
	\caption{Variation of the NAP-XPS intensities for (a) C\,$1s$, (b) O\,$1s$, and (c) Ni\,$2p$ measured during exposure of the gr/$20$\,ML-Ni/Ir(111) system to O$_2$ at fixed pressure ($0.1$\,mar) and fixed sample temperature ($150^\circ$\,C). The respective exposure time for every spectra is marked in the figure legend. All XPS spectra are shifted for clarity. The smoothed line through the experimental points is shown for every spectra.}
	\label{grOthickNiIr111_NAP_XPS}
\end{figure}

In order to study the dynamics of the oxygen intercalation process we performed time-dependent NAP-XPS experiments for different gr/$n$\,ML-Ni/Ir(111) systems and these results are presented for $0.5$\,ML-Ni (Fig.~\ref{grO05MLNiIr111_NAP_XPS}), $1.2$\,ML-Ni (Fig.~\ref{grO12MLNiIr111_NAP_XPS}), and $20$\,ML-Ni (Fig.~\ref{grOthickNiIr111_NAP_XPS}). First two experiments were performed in the same experimental conditions for the oxygen intercalation ($0.1$\,mbar and $150^\circ$\,C) and for the thick Ni layer we chose higher temperature ($250^\circ$\,C) in order to vary one of the parameters in comparison with the experimental conditions for thin Ni layers. 

For the gr/$0.5$\,ML-Ni/Ir(111) system, two C\,$1s$ components can be identified in the spectra after its fit (Fig.~\ref{grO05MLNiIr111_NAP_XPS}(a); fit components are shown for the bottom spectra as thin solid lines) -- at $284.9\pm0.1$\,eV and $284.2\pm0.1$\,eV, which can be assigned to the graphene covered areas with and without Ni intercalated, respectively. The intercalation of oxygen in this system on the initial stage (time between $0$\,min and $20$\,min) leads to the diffusion of oxygen atoms towards Ni islands and to the predominant oxidation of the Ni layer as indicated by the reduction of the intensity of the corresponding C\,$1s$ component. After this process is nearly complete, the areas where Ni was not intercalated start to fill up with oxygen atoms. The second step is indicated as a gradual shift of the centre of gravity of the whole C\,$1s$ peak towards lower binding energy and as a simultaneous reduction of its FWHM. Finally this peak has a single component indicating the complete oxygen layer intercalation and that in all areas (above Ir and above formed NiO) graphene layer is in contact with oxygen atoms. The measured at the same time O\,$1s$ and Ni\,$2p$ spectra (Fig.~\ref{grO05MLNiIr111_NAP_XPS}(b,c)) show the increase of the oxygen signal and complete oxidation of the thin Ni layer already at the initial steps of oxygen intercalation.

In case of gr/$1.2$\,ML-Ni/Ir(111), as was discussed earlier, the C\,$1s$ peak consists of the two components which can be assigned to two distinct areas for the strongly buckled graphene layer in this system: the higher and lower binding energy components are due to the larger and smaller interaction strengths of carbon atoms with Ni atoms in a lattice mismatched graphene layer~\cite{Pacile:2013jc}. The oxygen intercalation in this system leads to the gradual decrease of intensities of both gr-Ni-related C\,$1s$ components with the simultaneous growth of the intensity of the gr-NiO-related component (Fig.~\ref{grO12MLNiIr111_NAP_XPS}(a)). These changes indicate that during oxygen-intercalation the oxygen atoms open the boundary between graphene and metal step by step, and finally complete the intercalation with the full oxidation of the Ni layer and its conversion to NiO. This picture is also supported by the changes of intensities of the O\,$1s$ and Ni\,$2p$ lines (Fig.~\ref{grO12MLNiIr111_NAP_XPS}(b,c)) -- the intensity of the correlation satellite ($6$\,eV below the main line) slowly decreases with the simultaneous growth of the components which can be assigned to the formation of NiO. The behaviour of the shape of the Ni\,$2p$ emission line for the thin NiO layers which are formed during oxygen intercalation in gr/$0.5$\,ML-Ni/Ir(111) and gr/$1.2$\,ML-Ni/Ir(111) is similar to the one observed in the previous experiments on the growth of thin NiO on metallic surfaces, like Ag(001)~\cite{Caffio:2004hg} and Cu(111)~\cite{SnchezAgudo:2000cm}, where suppression of the shake-up satellite was observed without big energy shift of the main emission line. In the present experiments the energy shift of the Ni\,$2p_{3/2}$ peak (and the corresponding spin-orbit split counterpart) of about $300$\,meV to higher binding energies is observed. From the other side the absence of the big energy shift of the Ni\,$2p$ line can be assigned to the adsorbing of oxygen above Ni under graphene. However, further experiments and theoretical considerations are required to shed light on this effect.

The time dependent NAP-XPS experiments on oxygen intercalation in the gr/$20$\,ML-Ni/Ir(111) system show the similar intercalation behaviour like for gr/bulk-Ni(111) (Fig.~\ref{grOthickNiIr111_NAP_XPS}). In this case the single-component C\,$1s$ peak at $284.9\pm0.1$\,eV is slowly shifted to the smaller binding energies and reaches $284.1\pm0.1$\,eV after the intercalation process is complete (Fig.~\ref{grOthickNiIr111_NAP_XPS}(a)). At this stage the process of the oxygen intercalation and formation of the interface NiO layer saturates that is indicated by the stop of the change of the position of the C\,$1s$ line as well as by the stop of the growth of the intensity of the O\,$1s$ line and the respective components in the Ni\,$2p$ spectra (Fig.~\ref{grOthickNiIr111_NAP_XPS}(b,c)). It is interesting to note that in the O\,$1s$ spectra for this system the component which is assigned to the presence of OH$^{-}$ groups on the surface or can be connected with defects in the oxide layer appears already on the initial steps of the oxygen intercalation in gr/$20$\,ML-Ni/Ir(111). The slow decrease of the fraction of this component in the total intensity of the O\,$1s$ line as a function of time can be assigned with a small reduction of the number of defects in the NiO layer. However, comparison of the respective O\,$1s$ spectra for all three systems shows that number of defects in the formed NiO layer is much smaller in case of pre-intercalated thin Ni layers in gr/Ir(111).

\subsection{Attempts of the nitrogen intercalation in gr/Ir(111) and gr/Ni(111): NAP-XPS temperature dependence}

It is well known that adsorption of graphene on metallic or semiconducting surfaces always lead to the doping of graphene with its charge neutrality point placed above or below the Fermi level followed by the modification its electronic structure~\cite{Dedkov:2015kp,Batzill:2012,Tesch:2018hm}. Decoupling of graphene from the substrate aims to have a neutral graphene with high charge mobility. Different approaches on the intercalation of the atoms of, e.\,g., Ge, O, or F, partially succeed in this strategy~\cite{Verbitskiy:2015kq} or lead to the strongly $p$-doped graphene for the highly electronegative atoms~\cite{Dedkov:2017jn,Larciprete:2012aaa,Granas:2012cf,Walter:2011gf}. One of the possibilities to reach the charge neutrality of graphene is its decoupling from the substrate by nitrogen atoms~\cite{Caffrey:2015et}. However, up to now our knowledge, there are no studies on the intercalation of nitrogen in the gr/substrate interfaces. Some rare studies on the N-plasma or atomic-N treatment of such interfaces can be found in the literature~\cite{Masuda:2015eg,VelezFort:2014kh,Tsai:2014gf,Chai:2008fc} that leads to the interfaces nitridation or to the N-atom incorporation in a graphene layer.

\begin{figure}[t]
	\centering
	\includegraphics[width=0.5\columnwidth]{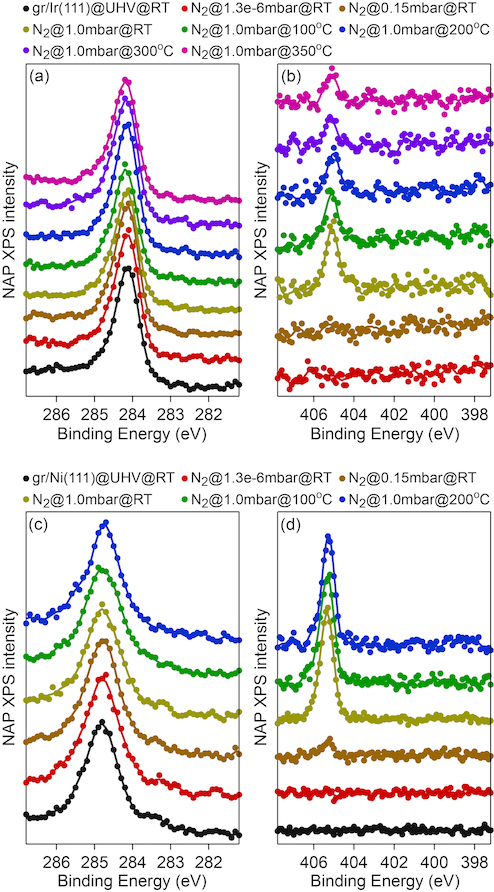}
	\caption{NAP-XPS intensities of (a,c) C\,$1s$ and (b,d) N\,$1s$ measured during exposure of gr/Ir(111) and gr/Ni(111), respectively, to N$_2$ at different gas pressures and temperatures of the substrate. All XPS spectra are shifted for clarity. The smoothed line through the experimental points is shown for every spectra.}
	\label{NAPXPS_N2_grIr111_grNi111}
\end{figure}

We have chosen the different approach and performed several attempts to intercalate nitrogen in gr/Ir(111) and gr/Ni(111) at high partial pressure of the N$_2$ gas and at the elevated substrate temperature. Fig.~\ref{NAPXPS_N2_grIr111_grNi111} shows C\,$1s$ (a and c) and N\,$1s$ (b and d) spectra for gr/Ir(111) and gr/Ni(111), respectively, during exposure of these systems to N$_2$ at different gas pressures and temperatures. One can clearly see that in both cases the C\,$1s$ line does not change during the sample treatment procedure -- no binding energy nor intensity changes, indicating the absence of the nitrogen intercalation. The same is true if N\,$1s$ spectra are considered -- the binding energy of this line is $405.1\pm0.1$\,eV and $405.3\pm0.1$\,eV for gr/Ir(111) and gr/Ni(111), respectively, that is characterised for N$_2$ molecules~\cite{Kramberger:2013dt}. (The slight difference in the binding energy for the N\,$1s$ line can be assigned to the different interaction strength at the respective gr-metal interface that can change the graphene-adsorbate interaction~\cite{Huttmann:2015hb}.) Moreover, this N\,$1s$ line disappear after gas is fully pumped from the UHV chamber, again supporting our conclusion on the absence of the nitrogen intercalation in the studied gr-metal interfaces. Our STM data for gr/Ir(111) and gr/Ni(111) (not shown here) collected before and after exposure these surfaces to N$_2$ also do not demonstrate any changes in the sample morphology.

\section{Conclusions}

We performed systematic NAP-XPS and STM studies of oxygen and nitrogen intercalation in the gr/Ni/Ir(111) interfaces of different morphologies which can be realized via pre-intercalation of different amounts of Ni. It is found that oxygen can be easily intercalated in all interfaces at the elevated temperature of the substrate and that intercalation is associated with the O$_2$-molecules dissociation. The intercalation of oxygen leads to the effective oxidation of the underlying Ni layer with the formation of the NiO layer of different thickness and morphology. This process leads to the effective decoupling of graphene in all cases with the resulting its strong $p$-doping. Our attempts to intercalate nitrogen at different conditions in gr/Ir(111) and gr/Ni(111) do not lead to the gas intercalation pointing the importance of the gas molecules dissociation at the gr-metal interface. Considering all available data on the gas intercalation in the gr-metal interfaces we can conclude that the bonding energy of atoms in the molecules and the probability to split in atoms at the metallic surface play crucial roles in the process. We suggest that the respective theoretical model has to be built which will allow to rationalise the studies of the systems on the intercalation of different gaseous species in different gr-metal and gr-semiconductor interfaces.

\section*{Experimental}
\label{ExpDet}

All experiments were performed in a customized station from SPECS Surface Nano Analysis GmbH for near-ambient pressure (NAP) spectroscopy and microscopy experiments consisting of three ultrahigh vacuum (UHV) chambers: NAP-XPS chamber, NAP-STM chamber, and molecular beam epitaxy (MBE) chamber, allowing for a sample transfer without breaking of the vacuum conditions. The background pressure of all three chambers is below than $< 3 \times 10^{-10}$\,mbar.

The SPECS Aarhus NAP-STM is equipped with a high-pressure reaction cell inside, which is sealed via a Viton$^{\textregistered}$ O-ring when performing high pressure experiments in the range of $10^{-5}$\,mbar -- $100$\,mbar. The maximum sample temperature during scanning can reach up to $770$\,K in UHV and about $500$\,K under near ambient pressure conditions and heating is performed by a halogen lamp heater which is mounted directly on the lid. 

The NAP-XPS chamber in the backfilling configuration is equipped with a SPECS PHOIBOS\,150 NAP hemispherical analyser (with $5$ stages of the differential pumping system) and monochromatic SPECS $\mu$-FOCUS\,600 NAP x-ray source: Al\,K\,$\alpha$ line ($h\nu=1486.6$\,eV), x-ray spot size on the sample $<250\mu\mathrm{m}$. This system allows NAP-XPS samples investigations in the pressure range for gasses up to $25$\,mbar with sample temperatures up to $1400$\,K by laser heating.

As a substrate for our experiments we used Ir(111) and Ni(111) crystals (MaTecK GmbH). Prior to every experimental run, these crystals were cleaned via cycles of Ar$^+$ sputtering ($2$\,kV, $5 \times 10^{-6}$\,mbar, $15$\,min) at room temperature with subsequent oxygen treatment at $700$\,K ($0.1$\,mbar, $10$\,min), followed by ultrahigh vacuum temperature flash for $3$\,min at $1800$\,K for Ir(111) and at $1000$\,K for Ni(111), respectively. Temperature for the graphene growth on Ir(111) or Ni(111) was estimated from the power vs. temperature curve (measured independently) or was measured by a chromel-alumel thermocouple, respectively. No subsurface Ar gas bubbles were observed for metallic surfaces used in our experiments as deduced from the STM and XPS measurements. Graphene overlayers were grown on Ir(111) and Ni(111) surface via chemical vapour deposition (CVD) procedure using C$_2$H$_4$ as described in the literature~\cite{Voloshina:2013dq,Dedkov:2017jn}, which lead to the graphene layers of very high quality as was confirmed by STM and XPS experiments. Thin intercalation Ni layers under graphene on Ir(111) were formed via annealing of the respective Ni/graphene/Ir(111) system at $800$\,K and the thickness of the intercalated Ni layer was estimated from the subsequent STM and XPS experiments. Intercalation of gases, oxygen and nitrogen, was studied via direct backfilling of the NAP-XPS chamber through the leak valves with the raised sample temperature, which was measured by a chromel-alumel thermocouple or by an infrared pyrometer. XPS spectra were measured directly during studies of the intercalation process in the ``live'' mode. The positions of the individual components in some spectra were determined after fitting procedure - single components were considered as a convolution of the Lorentzian and Doniach-Sunjic line shapes with the linear background.



\clearpage

\clearpage

\clearpage

\clearpage

\clearpage

\clearpage

\clearpage

\clearpage


\clearpage
\begin{figure}[t]
	\centering
	\includegraphics[width=0.5\columnwidth]{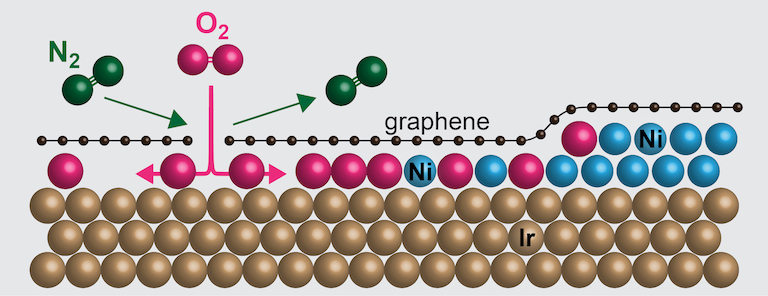}
\end{figure}

\end{document}